%% file: 00_main.tex
\documentclass[10pt, conference]{IEEEtran}
\IEEEoverridecommandlockouts
\usepackage[noadjust]{cite}
\usepackage{amsmath,amssymb,amsfonts}
\usepackage{algorithmic}
\usepackage{graphicx}
\usepackage{textcomp}
\usepackage{xcolor}
\def\BibTeX{{\rm B\kern-.05em{\sc i\kern-.025em b}\kern-.08em
    T\kern-.1667em\lower.7ex\hbox{E}\kern-.125emX}}
\begin{document}

\title{Characterizing relationships between primary miners in Ethereum by analyzing on-chain transactions}

\author{\IEEEauthorblockN{1\textsuperscript{st} Daniel Rincon Silva\thanks{This work was conducted under a research grant from Ripple's University Blockchain Research Innitiative (UBRI). }}
\IEEEauthorblockA{\textit{School of Information} \\
\textit{University of California Berkeley}\\
Berkeley, USA \\
dsrincon@berkeley.edu}

}

\IEEEoverridecommandlockouts
\IEEEpubid{\makebox[\columnwidth]{978-1-7281-7091-6/20/\$31.00~\copyright2020 IEEE \hfill} \hspace{\columnsep}\makebox[\columnwidth]{ }}

\maketitle
\IEEEpubidadjcol
\IEEEpubidadjcol

\begin{abstract}
It is widely accepted that Ethereum mining is highly centralized. Nonetheless, centralization has been mostly characterized by exclusively looking at the influence that independent miners or mining pools can have over the network. Moreover, models of mining behavior assume that miners are either unrelated or only relate via mining pools under highly structured and transparent agreements. If these assumptions and the predictions they entail were to be completely accurate, there would not be any evidence of on-chain transactions between miners, other than the ones expected from mining pool payouts. By looking at on-chain transactions between miners in the Ethereum Network we find that aside from the payouts from mining pools to small miners, there are also transactions that define relationships between mining pools, independent miners and between independent miners and mining pools. Furthermore, by characterizing the topology of the network of miner transactions, we find the emergence of highly connected clusters that control significant amounts of hashing power and exhibit relationships in the opposite direction of what theoretical models predict. This more nuanced characterization of mining centralization can help identify network vulnerabilities and inform protocol redesigns.

 
 

\end{abstract}

\begin{IEEEkeywords}
blockchain, ethereum, network analysis
\end{IEEEkeywords}

\input{01_introduction}
\input{02_related_work}
\input{03_methodology}
\input{04_concentration}
\input{05_network}

\input{06_components}
\input{07_control}
\input{08_discussion}

\input{09_conclusion}

\section*{Acknowledgment}

The author would like to thank Professor Joshua Blumenstock at the School of Information at UC Berkeley for his valuable comments.

\bibliography{bibliography.bib}
\bibliographystyle{IEEEtran}

\end{document}

%% file: 01_introduction.tex
\section{Introduction}

\subsection{Evolution of mining centralization}
Blockchain technology was first introduced with the Bitcoin network as a way to allow transactions between individuals without the need for trusted intermediaries to validate or store funds \cite{nakamoto2012bitcoin}. To guarantee this, it relies on public key cryptography to enable users to store and spend funds safely. Additionally, it introduced a decentralized Proof of Work (PoW) mechanism where independent validators (miners) compete for the right to validate transactions by solving a computational challenge.  They then follow a protocol to reach a consensus on the validated set of transactions, and the miner that performed the validation is rewarded with a fee. 

This trustless and permissionless model was intended to incentivize anyone to become a miner and create a system that was not dependent on a small number of centralized institutions, with the hope of making it more resilient and inclusive. Nonetheless, the specific design of the PoW mechanism quickly lead to a specialization in mining where only people with access to specialized hardware and low electricity costs could profitably participate in the mining process. By 2016 8 large Bitcoin miners controlled about 90\% of the mining (hashing) power in the network \cite{Bitonic_2016}. 

With the hope of building on top of Bitcoin's design and improving many of its features, including making it more decentralized, the Ethereum Network was launched in 2015. To avoid concentration, Ethereum designers tried to democratize access to mining by introducing a hashing algorithm that made it more challenging to produce specialized hardware to mine \cite{buterin2014ethereum}. Although this allowed independent small miners to participate in higher proportions, other forms of mining consolidation emerged. In particular, given that smaller miners had less chance of winning the computational challenge, it was in their interest to find ways to mitigate their risk by forming \textit{mining pools}. Mining pools aggregate miners that contribute hashing power to find the solution to the computational challenge. Once they collectively find a solution to the computational challenge, they share the fee proportionally to the amount of hashing power each miner contributed. Albeit mining pools are made up of independent miners, they are centrally operated and, in many ways, can be considered as a single miner that concentrates substantial hashing power \cite{luu2017smartpool}. 

\subsection{Issues with mining centralization}
The concentration of mining power is problematic, not only because it undermines the values of equal and open access that blockchain systems promote but also because it makes them more susceptible to random failures, targeted attacks, and misbehavior from bad actors. In particular, a high concentration of mining power can allow miners to manipulate the system to double-spend funds \cite{Jenkinson.2019}; selfish-mine to earn disproportionate rewards \cite{eyal2014majority};  delay transactions and take advantage of smart contracts \cite{flashboys}; or even censor specific transactions \cite{zcash_censor}. Some attacks like double-spending \cite{Jenkinson.2019} require the control of 51\% of the mining power, but others can be performed with less than 20\% \cite{eyal2014majority}. 

\subsection{Relationships between miners}
Disproportionate control of mining power (centralization) does not exclusively take the form of large miners or mining pools controlling significant amounts of hashing power; coalitions of miners, mining pools, or a combination of both can combine their hashing power to collude and potentially undermine Blockchain-based systems in the ways mentioned above. Coalitions of this sort can be permanent or ad-hoc and are probably a more pressing threat than the straightforward abuse of power that might be exerted by single large miners or mining pools. Vitalik Buterin, one of the designers behind Ethereum, stated in 2017 that Blockchain protocols should be designed  "in such a way as to reduce the incentive for validators/miners to engage in one-to-one 'special relationships'"\cite{buterin_2017_decentralization}.

Despite miner relationships being a pressing concern for protocol designers, these have been understudied in academic literature. Existing analysis assume that miners are either unrelated or only relate via mining pools with a well-defined and transparent structure. If these assumptions and the predictions they entail were to be completely accurate, there would not be any evidence of on-chain transactions between miners, other than the ones expected from mining pool payouts. This paper contributes to the existing literature by analyzing on-chain transactions between miners and uncovers economic relationships that are neither assumed or predicted by existing mining models. By applying network analysis in the way it has been utilized to examine banking systems, we characterize economic relationships between miners and provide:

 \begin{itemize}
  \item Key topological metrics of the network formed by transactions between miners.
  \item Identification of the roles assumed by different miners in this network.
  \item Description of clusters with highly connected miners.
  \item Identification of implicit dependencies derived from these relationships and the vulnerabilities they might create.
  
\end{itemize}

We hope our characterization can help researchers formulate better models of mining behavior and inform improvements in PoW protocols. 

This paper is divided into nine sections. The current section introduced our research problem and highlighted our specific contributions; Section 2 surveys related work; Section 3 presents our methodology; Sections 4 to 7 present our main results; sections 8 and 9 offer a discussion and present some conclusions. 

%% file: 02_related_work.tex
\section{Related Work}
\subsection{Network analysis in Blockchain}

Existing network analysis of Blockchain systems has mainly focused on analyzing fundamental properties of the overall transaction network, de-anonymizing entities, or tracking malicious activity.   Initial approaches like \cite{ron2012transgraph,baumann14exploring} perform statistical analysis summarizing the transaction history of the network in its initial four years. Some more recent work \cite{maesa17datadrivbit,kondor14richer} has extended this analysis using a larger dataset and leveraging on better frameworks and computation capabilities to calculate other properties of the transaction graph and their evolution. Research has also focused on analyzing the evolution of transactions by comparing transaction data with external data (i.e., geographic, pricing, entity labels, forum sentiment analysis) \cite{lischke16fouryears,mcginn18opendata} and showing that transaction behavior is consistent with traditional monetary theory. As for community analysis, the most recent literature focuses on applying Machine Learning techniques to cluster addresses into users or assign them to specific classes (e.g., exchanges, gambling, mining) \cite{monaco15transactionbeh,jourdan18entities,harlev18breaking}. Some other literature concerned with tracking malicious activity uses specific clustering heuristics to follow the transactions of specific accounts \cite{spagnulo14intelligence,meiklejohn16fistful}. 

More recent approaches have explored the network structure of ERC20 tokens in Ethereum \cite{somin18socialerc20} and compared it to that of social networks, while other work has analyzed the transaction activity associated with different types of smart contracts \cite{oliva2020exploratory}. 

\subsection{Understanding the mining ecosystem}

Existing literature has attempted to understand blockchain mining ecosystems, both empirically and theoretically. On the empirical front, the approaches have been limited to either quantifying the mining power controlled by different individual miners or pools or by looking at the activity of the Peer-2-Peer (P2P) network that connects miners. Initial approaches by \cite{gervais2014bitcoin, beikverdi2015trend} calculate the number of Bitcoins mined by different mining operations to estimate the hashing power controlled by each of them and effectively conclude that Bitcoin mining is centralized in very few miners and mining pools. 

More recent approaches, such as \cite{gencer2018decentralization,mariem2020all}, have analyzed the underlying P2P network that connects both Bitcoin and Etherum miners to evaluate its activity, the geographical location of different nodes and their level of connectivity. They conclude that mining operations are highly centralized in China with significant traffic activity originating in the US and EU and that albeit Etherum is more geographically dispersed, both miner networks exhibit a high degree of geographical clustering and concentration of mining power. 

Theoretical approaches to understanding the structure of mining ecosystems have consistently taken a dichotomous approach that tries to explain the emergence of mining pools from independent miners. The models offered by \cite{alsabah2019pitfalls,leonardos2020oceanic} show that given certain starting conditions and structural technology parameters, mining ecosystems will tend to evolve from independent miners to coalitions of mining pools. Similarly, \cite{arnosti2018bitcoin} models Bitcoin mining as a market with oligopolistic competition where cost asymmetries can lead to a small number of participants to control most of the mining power. Finally \cite{cong2019decentralized} models mining pools as mechanisms for risk-averse miners to share risks and shows that given specific initial conditions in the ability of pools to capture loyal, independent miners, mining pools will grow, but there will be endogenous mechanisms that limit their growth.

\subsection{Intermediary relationships in financial networks}

Although there is no consensus among regulators about how to categorize the role of miners, some researchers argue that they play a role akin to that of banks or other intermediaries in traditional financial systems. Under this interpretation, it is useful to see how the financial literature has explored intermediary relationships in financial systems. \cite{bech2010} studied the topology of the Federal Funds market in the US and found that relationships between banks are predictive of the way they specialize as either primarily borrowers or lenders and the interest rate they charge. 

More recent work has focused on understanding specific network properties of interbank systems, in particular, the extent to which the system can be controlled by influencing a subset of banks. In this line of work, \cite{delpini2013} explored the controllability of a section of the European interbank lending system and found that if the network is analyzed at large enough time scale, the percentage of nodes required to drive (control) the system is less than 10\%. They also found that the system's potential drivers tend not to be the largest lenders or those that have most connections. 

As shown above, existing attempts to understand mining ecosystems are limited to analyzing very coarse relationships (e.g., geographical location) or highly structured and limited ones (e.g., mining pool agreements). We believe that by applying some of the network analysis methods that have been used to study financial networks, we will be able to uncover more functional relationships between Ethereum miners to understand the structure of the ecosystem better. 

%% file: 03_methodology.tex
\section{Methodology}

To uncover economic relationships between miners, we parsed on-chain transaction data to extract transactions that only involved miners and defined a directed graph.  A node in the graph represented a miner, and weighted directed edges were defined between two nodes if there had been transactions going from one to the other. We then used this graph to calculate centralization properties, identify relationship clusters, and analyze the flow of transactions between miners. We present the data and formal graph definitions below. 

\subsection{Data}
To explore transactions between miners in the Ethereum Network, we used the dataset hosted in Google's Big Query platform, which parses data from the Ethereum blockchain and offers a SQL interface to explore block and transaction history. We queried all 8,863,264 blocks generated between the launch of the system on July 30, 2015, and November 3, 2019, to obtain all the addresses of miners that had mined those blocks and found that in this period 4,895 different addresses were involved in mining blocks. We then queried all 574,423,867 transactions within these blocks to extract those transactions where both the sender and receiver addresses corresponded to miners that had validated blocks in this same period. The method above led us to extract 292,882 transactions where only miners where involved. 
In addition to the BigQuery data used to extract block and transaction information, we used a time series of the daily exchange rate between Ether (ETH) and US Dollars (USD) provided by Etherscan for the relevant study period \cite{etherscan_price}. We used the ETH to USD exchange rate to convert all the relevant transactions to their value in USD at the time the transaction happened to make the value between transactions more comparable. 

\subsection{The Miner Transaction Network}

Given the lack of exact dates on our dataset we analyze incremental slices $p_k$ of the 8,863,264 blocks with $k=1,...,51$, where every slice gets $\hat{B}$ blocks added, where $\hat{B}$ is the average number of blocks mined every 30 days. For each slice $p_k$ we looked at the miners that mined blocks in that period as well as the miner transactions that took place. For each period $p_k$, we define a directed graph $G_k=(V_k,E_k)$ where $V_k$ is the set of all miners (vertices) and $E_k$ is the set of all ordered tuples $(m_i,m_j)_k$ (edges) where $m_i,m_j$ are miner addresses and there has been at least one transaction from $m_i$ to $m_j$ in the blocks between the genesis block and the block $|p_k|$ . In addition, we define an attribute function $A((m_i,m_j)_k):E\to\mathbb{R}^{+0}\times{R}^{+0}$ that maps every edge in the graph $(m_i,m_j)_k$ to a tuple $(v_{ij}^k,n_{ij}^k)$ where $v_{ij}^k$ is the total value (in USD) of all transactions between $m_i$ and $m_j$ in the period $|p_k|$ and $n_{ij}^k$ is the total number of transactions between these two miners in that same period. We call these edges $(m_i,m_j)_k$ that summarize all the transactions in one direction between two miners an 'economic relationship' from miner $i$ to miner $j$. 

%% file: 04_concentration.tex
\section{Concentration in Ethereum Mining}

For the final aggregation of the network $k=51$, we found that the distribution of blocks mined per miner is highly unequal and follows a power law distribution. The number of blocks mined per miner has a mean of 1,778 but a median of 4. Furthermore, the distribution is highly unequal with the Gini Coefficient of 0.991, meaning that 1\% of the miners active in the network mined over 99\% of the blocks. 
To evaluate concentration from a different perspective, we calculated the Herfindahl–Hirschman Index (HHI), which is used to measure the concentration of a given industry and is calculated as follows:

\begin{displaymath}
  HHI=\sum_{i=1}^{n} s_i^2
\end{displaymath}

Where $s_i$ is the percentage market share of the $i$th company in the industry, an HHI very close to 0 represents a market with no concentration, while an HHI of 10,000 is a fully concentrated market with one monopolistic firm.  For the final period of analysis ($k=51$), we found the HHI to be 846 for the Ethereum mining ecosystem. This value is higher than the HHI for the US banking industry in 2018 (523-617) but lower than what US regulation considers to be a concentrated industry (above $1500$) \cite{banking_industry}. As Figure ~\ref{fig:concentrationdist} shows, after a slight hike in concentration in the first 5-10 periods of analysis, the concentration of mining in the network has maintained relatively stable levels.

\begin{figure}[h]

  \centering
  \includegraphics[width=\linewidth]{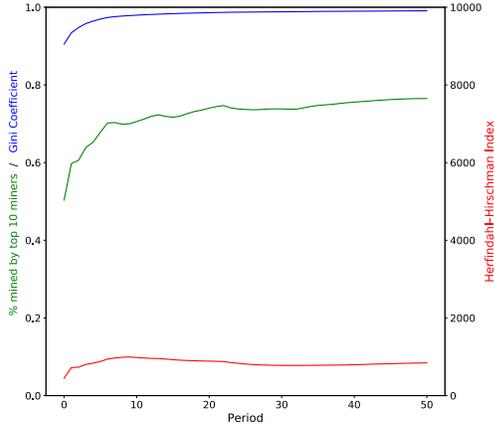}
  \caption{Evolution of mining concentration indicators.}
  \label{fig:concentration_dist}
\end{figure}

The top 10 mining addresses in the whole aggregate period mined 76.5\% of all blocks. According to the way addresses are identified by etherscan.io, all of the top 10 mining addresses correspond to mining pools (see Table ~\ref{tab:top10_mine}).

\begin{table}[htbp]

\caption{Percentage of Blocks mined by top 10 mining addresses in Ethereum between Jul 2015 and Nov 2019}
\begin{center}
\label{tab:top10_mine}
\begin{tabular}{|c|c|}

\hline
\textbf{Miner}& \textbf{Percentage of Blocks mined} \\

\hline
Ethermine & 18.9\%\\
    Nanopool & 10.9\%\\
    DwarfPool1 & 10.5\%\\
    Spark Pool & 9.7\%\\
    F2Pool & 9.0\%\\
    F2Pool: Old & 5.4\%\\
    MiningPoolHub & 5.0\%\\
    EthPool 2 & 3.1\%\\
    ethfans.org & 2.1\%\\
    Cointron 1 & 1.8\%\\
\hline
\end{tabular}
\label{tab1}
\end{center}
\end{table}

\begin{figure}[h]

  \centering
  \includegraphics[width=\linewidth]{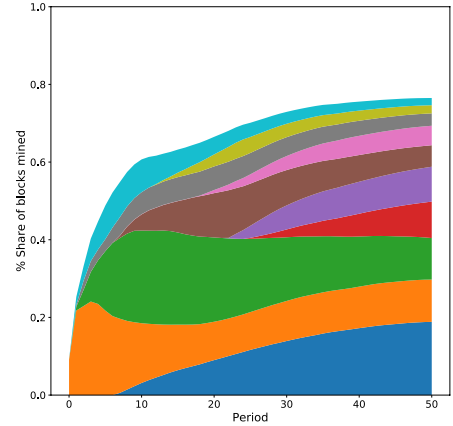}
  \caption{Evolution of  mining share for the top 10 miner addresses in the Ethereum network.}
  \label{fig:evol_top10}
\end{figure}

%% file: 05_network.tex
\section{Network of Transactions between Miners}

\subsection{Network Size}
When analyzing the final aggregate period for the miner transaction network ($k=51$), we found a total of $5,293$ economic relationships between miners. Out of the $4,895$ miners that mined blocks in that period, $2,899$ (representing 96\% of the historical hashing power) have at least one economic relationship with another miner. The density for this network, measured as the number of existing edges over the total number of possible edges, is 0.0213\%. Even though this density is lower than that of the Interbank Federal Funds market in the US, which is around 0.7\% \cite{bech2010}, it is at least $2,325$ times greater than the density of the transaction graph for all of the Ethereum Network\footnote{According to \cite{etherscan} and \cite{blockchair} by November 1, 2019, there were 79,001,882 unique addresses and 573,691,673 transactions in the Ethereum Network. Hence the density of the whole network by this time was $9.19\times10^{-6}$\% }. 

\begin{figure}[h]

  \centering
  \includegraphics[width=\linewidth]{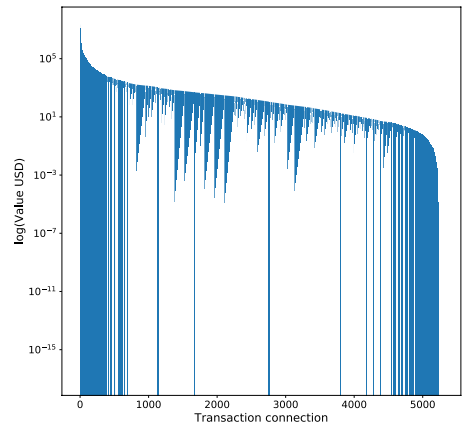}
  \caption{Economic relationships (transaction connections) ranked by transaction value}
  \label{fig:edge_dist}
\end{figure}

\begin{figure}[h]

  \centering
  \includegraphics[width=\linewidth]{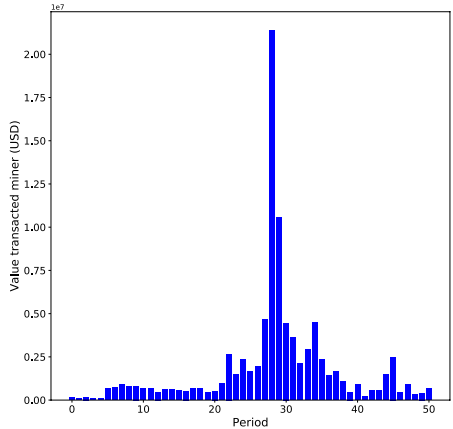}
  \caption{Value transacted between miners in tens of millions of USD for every period. Each period represents approximately 30 days of transactions. For example in the period with the highest amount of transactions (Period 28, between November and December 2017), the total value of transactions between miners was 21.4 Million USD. }
  \label{fig:evol_value}
\end{figure}

\begin{figure}[h]

  \centering
  \includegraphics[width=\linewidth]{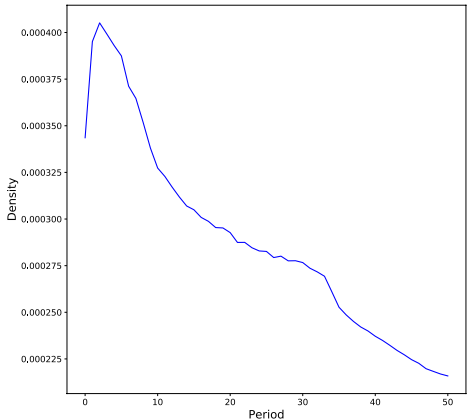}
  \caption{Evolution of the density of economic relationships.}
  \label{fig:evol_density}
\end{figure}

We found that the $5,293$ economic relationships established over the whole analysis period amount to $\$92,115,984$ US Dollars transacted between miners. As shown by Figure ~\ref{fig:edge_dist}, the distribution of value transacted in the economic relationships for the whole period, follows a power law distribution. The average value transacted in a relationship is of $\$17,403$, while the median is $\$136$. The Gini coefficient of this distribution is 0.98. The value transacted by relationships from or to the top 10 miners amounts to $\$70,925,800$ (see Table ~\ref{tab:top10_mine}), which is equivalent to $77\%$ of all the value transacted. 
The number of economic relationships rose sharply in the initial months of the network, but after that, the miner transaction network has become increasingly sparse, as shown by Figure ~\ref{fig:evol_density}.

\subsection{Degree distribution and correlation}

\begin{figure}[h]

  \centering
  \includegraphics[width=\linewidth]{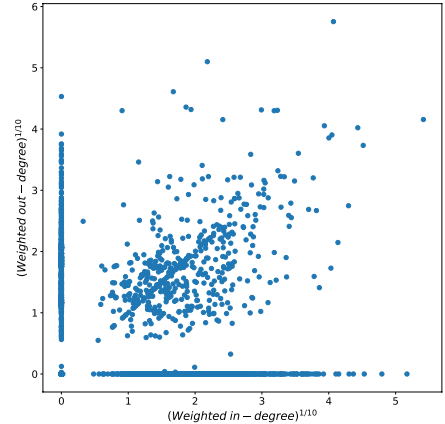}
  \caption{Correlation between in and out degrees of nodes in the network.}
  \label{fig:in_out_degree_cor}
\end{figure}

The weighted in-degree and out-degrees measures of a node in a directed network are defined as the number of incoming and outgoing edges to and from that node, multiplied (weighted) by the value of each edge. For any directed network, the average in-degree is equal to the average out-degree, and hence it is called the average \textit{degree}. As Figure ~\ref{fig:in_out_degree_cor} shows, both the distribution of the in and out degrees are highly skewed, while the average weighted degree is equal to $\$18,479$, the median weighted degree is $\$0$. Figure ~\ref{fig:in_out_degree_cor}  also reveals that there appears to be a marked specialization amongst miners where most nodes are either only receivers (out-degree = 0) or only senders (in-degree = 0), while a smaller group of miners engages in both sending and receiving transactions.

%% file: 06_components.tex
\section{Network components}

\subsection{Specialization in the Great Weakly Connected Component}
\begin{figure*}[h]

  \centering
  \includegraphics[width=12cm,height=6cm]{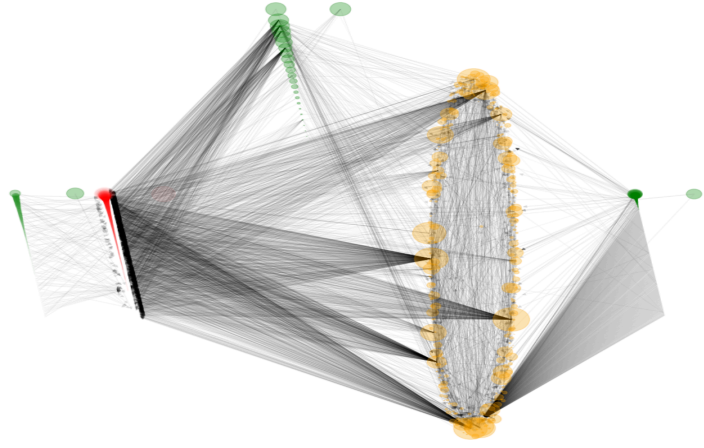}
  \caption{Great Weakly Connected Component. Green dots represent miners that only have outgoing transactions, yellow dots represent miners that have incoming and outgoing transactions, and red dots represent miners that only have incoming transactions. The size of the dot represents its (scaled) hashing power.}
  \label{fig:gwcc}
\end{figure*}

To better understand the topological features associated with the part of the network that has the most activity from the most relevant actors, we analyzed the structure of the Great Weakly Connected Component (GWCC). In a directed graph a weakly directed component is a maximal sub-graph where for every pair of vertices $m_i$, $m_j$ there is either a path from $m_i$ to $m_j$ or from $m_j$ to $m_i$. The GWCC is the largest of all the weakly connected components. For the period of study, we found that there are $2,244$ weakly connected components, but most of the network activity happens in the GWCC. The GWCC has $2,422$ miners that account for $91\%$ of the historical hashing power in the network. The value of the transactions in the GWWC is  $\$86,242,569$, which accounts for $94\%$ of all the transacted value. Since nodes out of the GWWC show isolated connections and do not represent a significant percentage of the transacted value, we focus our analysis on the GWCC. 
By highlighting specialized nodes in the GWCC, we can map out the relevant topological routes taken by transactions. Figure ~\ref{fig:gwcc} shows this detailed view of the GWCC with the three types of nodes: those that only send transactions (senders), those that only receive (receivers), and those that send and receive (mixed). The sender component has $876$ nodes that represent $4\%$ of the hashing power in the GWCC, the receiver component has $1,136$ nodes that represent $4.5\%$ of the hashing power in the GWCC, and the mixed component has $410$ nodes, that represent $91.5\%$ of the hashing power in the GWCC. 
The sender component has sent $\$4,107,145$ to the mixed component and $\$1,215,589$ to the receiver component, while the mixed component has sent $\$48,194,301$ and has internally transacted $\$32,725,534$.

\subsection{Topology of the Strongly Connected Components}

\begin{figure}[h]

  \centering
  \includegraphics[width=\linewidth]{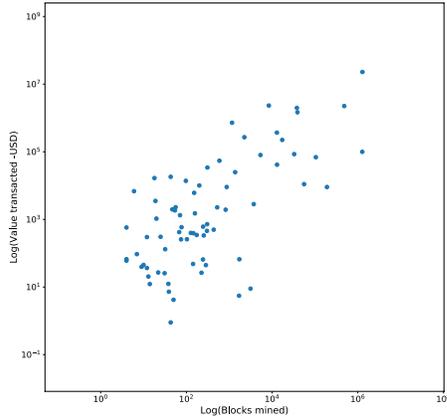}
  \caption{Hashing power vs. Internal value transacted in SCCs.}
  \label{fig:scc_hash_trans}
\end{figure}

To further characterize the topology of the network within the GWCC, we look at the sub-components in the mixed component, in particular at the Strongly Connected Components (SCCs). A Strongly Connected Component (SCC) is a sub-graph with a path that connects every pair of vertices. There are $73$ strongly connected components within the mixed component formed by $179$ different miners that represent $44\%$ of the hashing power in the GWCC. The mean and median size of an SCC is 2 miners, and the largest one has 9 miners and represents $16\%$ of the hashing power in the GWCC. SCCs have transacted $\$33,198,960$, and six of the top 10 miners belong to one SCC, with only two SCCs having two top 10 miners while the other four are in separate SCCs with no other miners from the top 10. There is a positive and highly significant positive correlation (Pearson r=0.69, p-value=$1.37\times10^{-11}$) between the hashing power of the miners in SCCs and the value they have transacted (see Figure ~\ref{fig:scc_hash_trans}).

%% file: 07_control.tex
\section{Control and Dominance}

\subsection{Transactions and hashing power hierarchy}

Given the dominance of mining pools in the Ethereum Network, one would expect that the transaction graph's topology would be mostly characterized by transactions from large mining pools to smaller miners representing the payouts from the former to the latter. In the previous section, we showed that this is not what is found when looking at the network's topology. In particular, Figure ~\ref{fig:gwcc} shows that there seems to be a significant amount of transactions that go against this expected pattern. To further quantify this behavior, we calculated how many of the economic relationships connected miners with other miners ranked higher than them in terms of hashing power. We call these relationships \textit{against the hierarchy}. We found that in the GWCC $1,088$ relationships that amounted to $\$10,578,921$ ($11\%$ of the total value transacted) were \textit{against the hierarchy}. On average, these relationships connected miners with other miners ranked $78\%$ higher in the hierarchy than them. 

\subsection{Driver nodes}

The fact that an economic relationship exists between miners can be indicative of many different things. One such possibility is that an economic relationship could create a potential for influence. In this section, we venture into exploring how vulnerable the system would be under the current topology if the existing economic relationships were used by some miners to exert a specific influence over others. 

To explore this possibility we first model this potential for influence as a canonical control problem over the transaction network. Say that the transaction network has linear time-invariant dynamics that can be modeled by:

\begin{displaymath}
  \frac{d\textbf{m}(t)}{dt}=A\textbf{m}(t)+B\textbf{u}(t)
\end{displaymath}

Where $\textbf{m}(t)=(m_1(t),...m_N(t))^T$ are the balances of the mining addresses of the system, $A$  is an $N\times N$ matrix that defines the strength of the economic relationships between miners and $B$ is an $N\times M$ ($M\leq N$) matrix that identifies which of the (driver) nodes are controlled by an external controller via a time-dependent vector $\textbf{u}(t)=(u_1(t),...u_M(t))^T$. \cite{liu2011controllability} showed that the minimum number of driver nodes needed to control the system described above is exactly the set of 'unmatched nodes' for a 'maximum matching' as long as all there are paths from the unmatched nodes to the matched ones. A 'maximum matching' is the maximal set of edges that do not share start or end nodes, while a node is said to be \textit{matched} if an edge in the 'maximum matching' points to it; otherwise, it is said to be \textit{unmatched}.
Our results yield that under this model and given the topology of the transaction network, $1,945$ nodes of the GWCC would need to be controlled ($47\%$ of the hashing power of the GWCC), in order to potentially have full control over it.

\subsection{Dominant sets with minimal hashing power}

Taking the previous section's exploration a step further, we model the potential for influence in a more direct way. We now try to understand how many (and which) nodes would be necessary to have direct influence over the whole network. More formally, we find a \textit{dominant} subset $D$ of nodes in the GWCC such that for every node in the GWCC not in $D$, there is a node in $D$ that has sent a transaction to it. By naively exploring different dominating sets starting from different points in the GWCC and using the algorithm presented in \cite{esfahanian2013connectivity}, we found the set composed of the nodes with the minimum hashing power needed to dominate all of the GWCC. This set is composed of $1,398$ nodes representing $25\%$ of the hashing power in the GWCC and only contains 3 miners from the top 10. 

We also explored the maximum vulnerability of the network under the current topology, assuming that miners could exert influence over other miners that were recipients of their transactions. We found that the set of nodes with the minimum hashing power needed to dominate a set with over $51\%$ ($51.07\%$) of the mining power is a set with only 7 nodes representing $0.88\%$ of the total hashing power. 

%% file: 08_discussion.tex
\section{Discussion}

\subsection{Evolution of the transaction network}

The miner transaction network exhibits a much higher density than expected from sampling a set with similar size from the overall Ethereum Network, which leads us to think that these relationships do have some practical significance. It is particularly interesting how large players have primarily driven this network.  Figure ~\ref{fig:evol_top10} shows how quickly the ecosystem consolidated, placing some miners at the top, indicating that shifts in power distribution are not the cause for the observed unexpected transaction patterns.  The volume of transactions in this market does not seem to be diminishing in absolute terms, as  Figure ~\ref{fig:evol_value} shows, yet it does seem to be decelerating. The existence of this network of transactions reveals that there are economic relationships beyond the simple miner-mining pool relationship assumed and predicted by previous work.  

\subsection{Transaction activity amongst large miners}

The particular structure of relationships between miners revealed by our analysis is indicative of underlying hierarchies and dependencies between different players. In particular, there seem to exist strongly connected groups that involve both independent miners and mining pools. These strong relationships are not explainable exclusively under models that assume that miners just come together in mining pools. It could be the case that some of these clusters of relationships and the roles of different miners within and between these clusters reveal fixed coalitions, ad-hoc alliances or different control structures (e.g., a single entity running multiple miner operations).  Further identifying the influence of these clusters and the implications of the underlying hierarchies on the reliability and efficiency of Ethereum can be an area for future work.

\subsection{Potential backdoors for control}

The advent of mining pools in PoW systems is argued by many, introduces decentralization because it allows more individuals to participate in the mining process. Nonetheless, mining pools centralize all of these resources under a central operator. This degree of control over the process can make mining pools as centralized as traditional large mining operations. The fact that large mining pools can potentially be subordinate to smaller players as suggested by some of the transaction analysis presented above should be a matter of concern. The nature of these asymmetric relationships should be better understood to evaluate the ways in which they could be exploited by malicious actors.

%% file: 09_conclusion.tex
\section{Conclusion}

By using network analysis to evaluate over 8 million blocks in the Ethereum network, the miners that validated them, and the transactions between them, we were able to characterize important patterns of how miners are economically related. Our approach had not been used in previous empirical work, and it revealed that relationships between miners are structured in ways that are not completely explained by existing theoretical models. Aside from the expected relationships between mining pools and small miners, we found transactions that reveal other types of coalitions and hierarchies between miners. We expect this characterization motivates further research into miner relationships and inspires potential protocol redesigns. 

%% file: 00_main.bbl
\begin{thebibliography}{10}
\providecommand{\url}[1]{#1}
\csname url@samestyle\endcsname
\providecommand{\newblock}{\relax}
\providecommand{\bibinfo}[2]{#2}
\providecommand{\BIBentrySTDinterwordspacing}{\spaceskip=0pt\relax}
\providecommand{\BIBentryALTinterwordstretchfactor}{4}
\providecommand{\BIBentryALTinterwordspacing}{\spaceskip=\fontdimen2\font plus
\BIBentryALTinterwordstretchfactor\fontdimen3\font minus
  \fontdimen4\font\relax}
\providecommand{\BIBforeignlanguage}[2]{{%
\expandafter\ifx\csname l@#1\endcsname\relax
\typeout{** WARNING: IEEEtran.bst: No hyphenation pattern has been}%
\typeout{** loaded for the language `#1'. Using the pattern for}%
\typeout{** the default language instead.}%
\else
\language=\csname l@#1\endcsname
\fi
#2}}
\providecommand{\BIBdecl}{\relax}
\BIBdecl

\bibitem{nakamoto2012bitcoin}
\BIBentryALTinterwordspacing
S.~Nakamoto, ``Bitcoin: A peer-to-peer electronic cash system,'' 2009.
  [Online]. Available: \url{http://www.bitcoin.org/bitcoin.pdf}
\BIBentrySTDinterwordspacing

\bibitem{Bitonic_2016}
\BIBentryALTinterwordspacing
Bitonic, ``A while ago part ii of the scaling bitcoin workshops took place and
  a lot has happened since then.'' Jun 2016. [Online]. Available:
  \url{https://medium.com/@Bitonicnl/a-while-ago-part-ii-of-the-scaling-bitcoin-workshops-took-place-and-a-lot-has-happened-since-then-f4b6f05bfa4c}
\BIBentrySTDinterwordspacing

\bibitem{buterin2014ethereum}
\BIBentryALTinterwordspacing
V.~Buterin, ``Ethereum: A next-generation smart contract and decentralized
  application platform,'' 2014, accessed: 2016-08-22. [Online]. Available:
  \url{https://github.com/ethereum/wiki/wiki/White-Paper}
\BIBentrySTDinterwordspacing

\bibitem{luu2017smartpool}
L.~Luu, Y.~Velner, J.~Teutsch, and P.~Saxena, ``Smartpool: Practical
  decentralized pooled mining,'' in \emph{26th $\{$USENIX$\}$ Security
  Symposium ($\{$USENIX$\}$ Security 17)}, 2017, pp. 1409--1426.

\bibitem{Jenkinson.2019}
\BIBentryALTinterwordspacing
G.~Jenkinson, ``Ethereum classic 51
  \emph{Cointelegraph}, 2019. [Online]. Available:
  \url{https://cointelegraph.com/news/ethereum-classic-51-attack-the-reality-of-proof-of-work}
\BIBentrySTDinterwordspacing

\bibitem{eyal2014majority}
I.~Eyal and E.~G. Sirer, ``Majority is not enough: Bitcoin mining is
  vulnerable,'' in \emph{International conference on financial cryptography and
  data security}.\hskip 1em plus 0.5em minus 0.4em\relax Springer, 2014, pp.
  436--454.

\bibitem{flashboys}
P.~Daian, S.~Goldfeder, T.~Kell, Y.~Li, X.~Zhao, I.~Bentov, L.~Breidenbach, and
  A.~Juels, ``Flash boys 2.0: Frontrunning, transaction reordering, and
  consensus instability in decentralized exchanges,'' 04 2019.

\bibitem{zcash_censor}
\BIBentryALTinterwordspacing
M.~Moss, ``Mining pool censorship could make zcash “mostly unusable”,''
  \emph{Cryptoslate}, 2019. [Online]. Available:
  \url{https://cryptoslate.com/mining-pool-censorship-zcash-unusable/}
\BIBentrySTDinterwordspacing

\bibitem{buterin_2017_decentralization}
\BIBentryALTinterwordspacing
V.~Buterin, ``The meaning of decentralization - vitalik buterin - medium,'' Feb
  2017. [Online]. Available:
  \url{https://medium.com/@VitalikButerin/the-meaning-of-decentralization-a0c92b76a274}
\BIBentrySTDinterwordspacing

\bibitem{ron2012transgraph}
D.~Ron and A.~Shamir, ``Quantitative analysis of the full bitcoin transaction
  graph,'' 11 2012.

\bibitem{baumann14exploring}
A.~Baumann, B.~Fabian, and M.~Lischke, ``Exploring the bitcoin network,''
  vol.~1, 04 2014.

\bibitem{maesa17datadrivbit}
D.~Maesa, A.~Marino, and L.~Ricci, ``Data-driven analysis of bitcoin
  properties: exploiting the users graph,'' \emph{International Journal of Data
  Science and Analytics}, 09 2017.

\bibitem{kondor14richer}
D.~Kondor, M.~Pósfai, I.~Csabai, and G.~Vattay, ``Do the rich get richer? an
  empirical analysis of the bitcoin transaction network,'' \emph{PloS one},
  vol.~9, p. e86197, 02 2014.

\bibitem{lischke16fouryears}
M.~Lischke and B.~Fabian, ``Analyzing the bitcoin network: The first four
  years,'' \emph{Future Internet}, vol.~8, 03 2016.

\bibitem{mcginn18opendata}
D.~McGinn, D.~McIlwraith, and Y.~Guo, ``Toward open data blockchain analytics:
  A bitcoin perspective,'' \emph{Royal Society Open Science}, vol.~5, 02 2018.

\bibitem{monaco15transactionbeh}
V.~Monaco, ``Identifying bitcoin users by transaction behavior,'' 04 2015.

\bibitem{jourdan18entities}
M.~Jourdan, S.~Blandin, L.~Wynter, and P.~Deshpande, ``Characterizing entities
  in the bitcoin blockchain,'' 11 2018, pp. 55--62.

\bibitem{harlev18breaking}
M.~Harlev, H.~Yin, K.~Langenheldt, R.~R. Mukkamala, and R.~Vatrapu, ``Breaking
  bad: De-anonymising entity types on the bitcoin blockchain using supervised
  machine learning,'' 01 2018.

\bibitem{spagnulo14intelligence}
M.~Spagnuolo, F.~Maggi, and S.~Zanero, ``Bitiodine: Extracting intelligence
  from the bitcoin network,'' vol. 8437, 03 2014, pp. 457--468.

\bibitem{meiklejohn16fistful}
S.~Meiklejohn, M.~Pomarole, G.~Jordan, K.~Levchenko, D.~Mccoy, G.~Voelker, and
  S.~Savage, ``A fistful of bitcoins: Characterizing payments among men with no
  names,'' \emph{Communications of the ACM}, vol.~59, 04 2016.

\bibitem{somin18socialerc20}
S.~Somin, G.~Gordon, and Y.~Altshuler, ``Social signals in the ethereum trading
  network,'' 05 2018.

\bibitem{oliva2020exploratory}
G.~A. Oliva, A.~E. Hassan, and Z.~M.~J. Jiang, ``An exploratory study of smart
  contracts in the ethereum blockchain platform,'' \emph{Empirical Software
  Engineering}, pp. 1--41, 2020.

\bibitem{gervais2014bitcoin}
A.~Gervais, G.~O. Karame, V.~Capkun, and S.~Capkun, ``Is bitcoin a
  decentralized currency?'' \emph{IEEE security \& privacy}, vol.~12, no.~3,
  pp. 54--60, 2014.

\bibitem{beikverdi2015trend}
A.~Beikverdi and J.~Song, ``Trend of centralization in bitcoin's distributed
  network,'' in \emph{2015 IEEE/ACIS 16th International Conference on Software
  Engineering, Artificial Intelligence, Networking and Parallel/Distributed
  Computing (SNPD)}.\hskip 1em plus 0.5em minus 0.4em\relax IEEE, 2015, pp.
  1--6.

\bibitem{gencer2018decentralization}
A.~E. Gencer, S.~Basu, I.~Eyal, R.~Van~Renesse, and E.~G. Sirer,
  ``Decentralization in bitcoin and ethereum networks,'' in \emph{International
  Conference on Financial Cryptography and Data Security}.\hskip 1em plus 0.5em
  minus 0.4em\relax Springer, 2018, pp. 439--457.

\bibitem{mariem2020all}
S.~B. Mariem, P.~Casas, M.~Romiti, B.~Donnet, R.~Stutz, and B.~Haslhofer, ``All
  that glitters is not bitcoin--unveiling the centralized nature of the btc
  (ip) network,'' \emph{arXiv preprint arXiv:2001.09105}, 2020.

\bibitem{alsabah2019pitfalls}
H.~Alsabah and A.~Capponi, ``Pitfalls of bitcoin’s proof-of-work: R\&d arms
  race and mining centralization,'' \emph{Available at SSRN 3273982}, 2019.

\bibitem{leonardos2020oceanic}
N.~Leonardos, S.~Leonardos, and G.~Piliouras, ``Oceanic games: Centralization
  risks and incentives in blockchain mining,'' in \emph{Mathematical Research
  for Blockchain Economy}.\hskip 1em plus 0.5em minus 0.4em\relax Springer,
  2020, pp. 183--199.

\bibitem{arnosti2018bitcoin}
N.~Arnosti and S.~M. Weinberg, ``Bitcoin: A natural oligopoly,'' \emph{arXiv
  preprint arXiv:1811.08572}, 2018.

\bibitem{cong2019decentralized}
L.~W. Cong, Z.~He, and J.~Li, ``Decentralized mining in centralized pools,''
  National Bureau of Economic Research, Tech. Rep., 2019.

\bibitem{bech2010}
M.~Bech and E.~Atalay, ``The topology of the federal funds market,''
  \emph{Physica A: Statistical Mechanics and its Applications}, vol. 389, pp.
  5223--5246, 11 2008.

\bibitem{delpini2013}
D.~Delpini, S.~Battiston, M.~Riccaboni, G.~Gabbi, F.~Pammolli, and
  G.~Caldarelli, ``Evolution of controllability in interbank networks,''
  \emph{Scientific reports}, vol.~3, p. 1626, 04 2013.

\bibitem{etherscan_price}
\BIBentryALTinterwordspacing
etherscan.io, ``Ether - usd exchange rate | etherscan,'' 2015, accessed:
  2019-11-01. [Online]. Available: \url{https://etherscan.io/chart/etherprice}
\BIBentrySTDinterwordspacing

\bibitem{banking_industry}
\BIBentryALTinterwordspacing
G.~Baer, ``The banking industry is unconcentrated, and will remain so after the
  bb\&t/suntrust merger,'' \emph{Banking Policy Institute}, 02 2019. [Online].
  Available:
  \url{https://bpi.com/the-bbt-corp-and-suntrust-banks-merger-will-not-raise-the-concentration-of-the-banking-industry/}
\BIBentrySTDinterwordspacing

\bibitem{etherscan}
\BIBentryALTinterwordspacing
etherscan.io, ``Ethereum unique addresses chart | etherscan,'' 2015, accessed:
  2019-11-01. [Online]. Available: \url{https://etherscan.io/chart/address}
\BIBentrySTDinterwordspacing

\bibitem{blockchair}
\BIBentryALTinterwordspacing
``Ethereum total transaction count chart blockchair 2020,'' 2019, accessed:
  2019-11-01. [Online]. Available:
  \url{https://blockchair.com/ethereum/charts/total-transaction-count}
\BIBentrySTDinterwordspacing

\bibitem{liu2011controllability}
Y.-Y. Liu, J.-J. Slotine, and A.-L. Barab{\'a}si, ``Controllability of complex
  networks,'' \emph{nature}, vol. 473, no. 7346, p. 167, 2011.

\bibitem{esfahanian2013connectivity}
A.-H. Esfahanian, ``Connectivity algorithms,'' in \emph{Topics in structural
  graph theory}.\hskip 1em plus 0.5em minus 0.4em\relax Cambridge University
  Press, 2013, pp. 268--281.

\end{thebibliography}
